# Evidence from the Special Relativity and Blackbody Radiation Theories for the Existence of Photons Possessing Zero Kinetic Energy

by

**Robert J. Buenker**[*]

FB C – Mathematik und Naturwissenschaften

Bergische Universität Wuppertal, Gaußstr. 20,

D-42119, Wuppertal Germany

**Abstract**

The traditional interpretation of radiative emission and absorption asserts that photons are created and annihilated in such processes. A *Gedanken* experiment is considered in which kinetic energy from observed photons is systematically removed until a limit of zero is reached. With the help of the relativistic Doppler effect it is shown that even for infinitesimally small kinetic energies the photons continue to exist, since in other inertial systems they will be observed to have a much higher energy/frequency falling in an easily detectable range. It is possible to formulate an alternative explanation for absorption and emission processes on this basis in terms of real photons with exactly zero kinetic energy being present before or after radiative interactions. Bolstering this hypothesis is the fact that the statistical mechanical treatment of photons interacting with oscillators in blackbody radiation theory predicts an infinite density of photons of this energy, both in the original Planck formulation employing Maxwell-Boltzmann statistics and in the subsequent Bose-Einstein description. These considerations demonstrate that the E = 0 state is greatly preferred as the product of absorption because of the requirement to have the interaction occur in a relatively narrow region of space- time. There is thus strong evidence that photons are not created and annihilated in radiative processes but simply have their kinetic energy changed either to or from a zero value. Accordingly a very high density of zero-energy photons is expected to exist uniformly throughout the universe. Finally, this development suggests that one should subject the creation-annihilation hypothesis to careful scrutiny in other areas of physics as well.

**I. Introduction**

The theory of the photon's role in processes involving the absorption or emission of electromagnetic radiation is based on the principle of creation and annihilation of matter. When a real photon interacts with an atom in an absorption process, it loses all of its kinetic energy and in view of its vanishing rest mass it is thought to lose its existence as well, *i.e.,* to be annihilated. Accordingly the concept of a rest mass for the photon is somewhat hypothetical in nature, since in this standard approach real photons must always possess a non-vanishing amount of kinetic energy and therefore move with the speed of light away from any observer. Yet in many other ways it has been found that single photons behave no less as particles than any of their counterparts with non-zero rest mass. With the help of the Bohr relation [1], $E=h\nu$, and de Broglie's law [2,3], $p=h/\lambda$, it has been possible to demonstrate that they satisfy ordinary conservation laws for energy, linear and angular momentum [3]. This emphasis on the particle or "corpuscular" nature of photons in the current century had its beginning with Einstein's explanation for the photoelectric effect [4], in which it became unavoidable to assume that single entities were responsible for the observed effects rather than a strictly delocalised form of matter.

The discovery that radio- and microwaves, infrared, visible and ultraviolet radiation, x-rays and gamma rays all differ only in the amount of kinetic energy of their individual constituents led to a significant unification in physical theory [5]. On this basis it was realised that photons can possess varying amounts of kinetic energy over a continuous range, exactly as other systems such as nuclei or electrons, which were long considered to be particles. Only the restriction that photons cannot exist at rest, distinguishes them from other conventional particles in this respect. Lowering the kinetic energy of an electron never causes it to be annihilated, for example, but rather ultimately brings it to a state of rest relative to the observer. Resistance to the idea of a further unification of the theory whereby photons and electrons would simply be taken to be particles of different rest mass, each of which is capable of taking on any amount of kinetic energy from zero upwards, is provided by conventional radiative absorption and emission experiments, however. After all, photons are only observed before an absorption process has occurred but not after, just as they can be detected at the conclusion of radiative emission but not before. Such observations support an interpretation in terms of the creation and annihilation of the detectable photons in these processes, which therefore opposes the view that these particles can exist with vanishing kinetic energy similarly as do electrons and other particles of non-zero rest mass. Also supporting the conclusion that photons are annihilated in radiative absorption processes is the fact that they invariably give up *all* their energy to atoms with which they interact, whereas electrons are known to only give up a portion of their kinetic energy in inelastic collisions with the same partners.

Yet if the creation-annihilation interpretation is made extremely plausible by such experience, it still needs to be recognised that there must always be an element of uncertainty in such a conclusion because there is no direct means of proving, for example, that a photon thought to be created did not in fact exist in an *undetectable* state prior to the radiative emission process. By definition a process involving creation or annihilation of matter is one in which either a portion of the reactants or a portion of the products is totally missing, viz. it is a reaction either to or from nothing. Since the domain of the physical sciences is restricted to observable phenomena, one is therefore left with a dilemma of sorts whenever such an interpretation is proposed, since there is clearly something about nothing that defies observation. Lack of detection can have several interpretations, only one of which is lack of existence of the object being sought, the other most common one being the inadequacy of available measuring devices for the purpose at hand. Experimentally it is impossible to distinguish between material particles that have ceased to exist and those that are simply lost from view for a period of time. In many ways it is ironic that much of the scientific progress of the previous two centuries can be attributed to the conscious decision to reject the premise that things simply disappear, whereas in more modern times exactly the opposite interpretation has been axiomatic.

Especially since the principle of creation and annihilation of matter plays such an integral role in the theory of subatomic physics, it seems essential to subject it to careful scrutiny before accepting it as a proven fact. In the present study an effort will be made to carry out such an analysis by employing the standard logical method of assuming that the converse of the above principle is true and then proceeding to look for a contradiction that demonstrates that this alternative position is untenable. Ultimately this amounts to posing the broader question of whether the body of experimental information alluded to above can be understood in a consistent manner which denies the occurrence of either particle annihilation or creation in any physical transformation as yet observed. Ideally one would like to find an experiment that is capable of providing a definitive result that clearly distinguishes the two possibilities. If there were no such experiment one might be tempted to say that there is a complete equivalence between the two theories, but it is clear at the outset that this would be an unsatisfactory conclusion because of the diametrically opposite premises on which they are based, namely whether material particles are created and annihilated or not.

**II. Conclusions Based on the Relativistic Doppler Effect**

The absorption and emission of electromagnetic radiation that accompanies a transition between two states of an atomic or molecular system is one of the simplest examples in modern physical theory in which the principle of creation and annihilation of matter is applied. Photons are said to be created in emission [eq. (1)] and destroyed in absorption [eq. (2)] processes.

$$\text{emission:} \quad A^* \rightarrow A + \gamma \, (E = h\nu) \tag{1}$$

$$\text{absorption:} \quad A + \gamma \, (E = h\nu) \rightarrow A^* \tag{2}$$

Experimentally one has evidence for the existence of the atomic system A on both sides of each of these equations, but the electromagnetic radiation is only detected either before or after the process has occurred. On this basis, it is easy to understand how the interpretation in terms of created or annihilated photons gained acceptance, but as mentioned in the Introduction, lack of detection of an object does not constitute proof of its nonexistence.

The theory of special relativity [6] adds substance to this discussion by pointing out that since the observed photons always move at the speed of light that their rest mass $m_0$ must be exactly zero. Because of the law of mass dilation,

$$m = m_0 \, (1 - v^2/c^2)^{-1/2} = m_0 \, (1 - \beta^2)^{-1/2} = m_0 \, ? \, , \tag{3}$$

such a particle can possess finite energy (or relativistic mass m) as long as v = c. If the photon's speed were to fall below this value, it follows from the same formula that its relativistic mass is the same as its (null) rest mass, from which it is generally concluded that it must therefore cease to exist. There is a potential flaw in the latter argument, however, because we do not really know that a particle of zero relativistic mass cannot continue to exist. What we do know is that according to the Bohr relation [1] the frequency ν associated with photons of this (zero) energy would be of *vanishing magnitude and therefore be undetectable by conventional spectroscopic techniques.* The Bohr relation thus tells us that, if there are such massless photons, there is no conceivable way of verifying this fact directly by purely experimental means.

One way of approaching the question of whether massless photons can exist is through the following *Gedanken* experiment. Consider a monoenergetic beam of photons that are in an easily detectable frequency range and imagine the effect of gradually withdrawing energy from them. At first one would just measure a different frequency for the photon beam and there would be no question of their continued existence. When all the energy is withdrawn, the possibility arises that the photons suddenly cease to exist, as the creation and annihilation hypothesis asserts, but the alternative conclusion is that they simply pass to an undetectable state by virtue of the zero frequency of the corresponding radiation. Since there are definite limits as to how long a wavelength or how short a frequency can be measured, a third possibility can also be envisioned, namely that even photons of non-zero energy might lose their existence when a point is reached for which experimental detection is unfeasible. The latter result would therefore also be consistent with the creation-annihilation hypothesis.

It might be thought that the above procedure is impractical because of the lack of a means of systematically withdrawing energy from the photon beam, but the theory of special relativity provides a suitable vehicle for overcoming this objection. Because of the Doppler effect it is possible to continuously vary the frequency of a beam of photons by simply changing the speed of the source relative to the observer. The observed frequency *ν* is related to that at the source *ν'* by the relation [7]:

$$\nu' = \gamma \nu (1 + \beta \cos ?) \tag{4}$$

where ? is the polar angle with respect to the direction of propagation. In principle the frequency of the photon beam can be lowered to any value down to (but not including) zero simply by increasing the speed of the source relative to the observer (transverse Doppler effect). Since it is clear that the existence or nonexistence of the photons in the beam is totally independent of how fast the observer is moving relative to the source, the third possibility indicated above cannot occur. In other words this argument shows definitively that *the energy/frequency of the photons can be lowered arbitrarily close to zero without having the particles themselves go out of existence.*

The same formula tells us something interesting about the possibility of massless photons existing as well, however. Since their frequency must vanish by the Bohr relation, $E = h\nu$, it follows that the measured frequency $\nu$ must also be zero regardless of the relative speed of the source to the observer (assuming $\gamma$ is finite, i.e., $\beta < 1$). If this were not the case it would be possible to rule out the existence of massless photons on this basis, since then one would predict that their associated frequency could be moved into a detectable range just by moving away from such a source at high speed. Since no such effect has ever been observed experimentally, this would be sufficient to vitiate the hypothesis of real (existing) massless photons, but once again it is seen that such a conclusion is without basis according to the theory of special relativity.

The latter point is interesting in another context, namely with regard to the definition of Minkowski spaces. The energy-momentum four-vector for a massless photon ($E = 0$) constitutes the identity in this linear space. Any particle with nonzero rest mass cannot be characterised by such a null vector. Since the definition of a linear space requires the existence of an identity, it can be argued that the failure to allow for real massless particles negates the possibility of regarding the linear manifold of physical energy four-vectors as a concretisation of this mathematical entity. One could attempt to save the situation by claiming that the *nonexistent* rest state of photons or other massless particles serves the function of an identity in this case, but since all other vectors in the same space are quite real there is at least an element of inconsistency in this approach. It is certainly more straightforward to admit the existence of real massless photons (and other particles with vanishing rest mass) and thereby fulfill all the conditions for a linear space without the need for clarifying physical interpretations. It is also noteworthy that the linear manifold of space-time coordinates does have a null vector that serves as the identity for this Minkowski space that requires no complicated interpretation. Thus the claim that the situation is more complex for energy-momentum four-vectors does not receive support on this basis either.

Returning to the *Gedanken* experiment based on the Doppler effect, one sees that there are only two possibilities open. The creation and annihilation of matter hypothesis must claim that as energy is withdrawn from the photon beam, the individual particles retain their existence continuously until the very moment that the limit of $E = 0$ is reached, at which point the entire system disappears. There is an alternative to this view which defies experimental contradiction, however, and that is that the photons are continuously in existence regardless of the amount of energy they possess, including $E = 0$, and that in this way they are no different than any other particles with non-zero rest mass, *i.e.,* they too can exist in a state of vanishing kinetic energy.

A clear analogy exists between this situation and the limit-taking procedure in mathematics. There, if it can be shown that a particular value $y_0$ for a function $f(x)$ can be approached as closely as desired simply by having the independent variable differ from some value $x_0$ by an arbitrarily small amount, it can be concluded that the limiting value of the function $f(x_0)$ exists at that point and that its numerical value is $y_0$. The latter definition is particularly useful when one is confronted with a situation in which $f(x_0)$ cannot be evaluated by conventional means, as for example for the ratio $\sin x / x$ for $x = 0$. In the present case of physical interest, it has been shown that regardless of how closely the observed frequency of the photon beam approaches zero, it is possible with the help of the Doppler effect to verify its existence. Just because we are prohibited from doing the same thing when its frequency vanishes exactly should not keep us from concluding that the correct result can be successfully inferred on the basis of findings obtained for a frequency only infinitesimally removed from that of the limiting case. Since this line of reasoning will be useful in subsequent discussions as well, it will be referred to on a general basis as *the principle of infinitesimally extended validity.* In the present case it simply means that since photons can be shown to exist for any (positive) frequency infinitesimally close to zero, it must be concluded that they also exist when their frequency (energy) exactly vanishes as well.

As a final remark in this section, it is worthwhile to pose a question about the properties of real massless photons, specifically at what speed they are allowed to travel relative to a given observer. Because of eq. (3) it has been concluded that photons of non-zero frequency must move at the speed of light c *in vacuo.* Examination of the same equation shows that *no such limit can be inferred for photons with vanishing relativistic mass* m, however. Any value of the relative speed v satisfies eq. (3) under these conditions, since one simply has a value of zero on both sides of the equation for all values of $\gamma$. This means that massless photons can move with the speed of light relative to the observer, but *they can also be at rest or move at any other relative speed smaller than c as well.* Since m = 0, the momentum p = mv will always be of vanishing magnitude for massless photons regardless of their speed v, and hence E = pc will remain zero as well. This result means that as the photon's energy changes from an infinitesimally small value to one of exactly vanishing magnitude, its relative speed may change in a continuous manner from its initial value of v = c down to and including a final value of v = 0 in a particular inertial system. It is not possible to confirm this result by direct experimental means, but just as before with the more basic question of whether real massless photons exist or not, it can be inferred by applying the principle of infinitesimally extended validity, in this case with the aid of eq. (3). In other words, a formula that always correctly predicts the speed of photons with infinitesimally small kinetic energy must also prove reliable for the corresponding limiting case in which they have exactly zero energy.

## III. Statistical Mechanics of Massless Photons: Conclusions Based on the Theory of Blackbody Radiation

Once the possibility has been grasped that massless photons exist, the question arises as to what their number density might be. The impossibility of detecting such particles because of the vanishing frequency associated with mono-energetic beams of them prevents one from designing an experimentally feasible counting scheme for this purpose, so it is again necessary to resort to indirect means to settle this issue. For this purpose it is instructive to go back to the very first experiment that was successfully interpreted on the basis of quantum mechanics, the phenomenon of blackbody radiation. Planck was able to provide a satisfactory explanation for the observed intensity distribution of a perfect absorber as a function of temperature by introducing a quantum hypothesis [8] into the classical theory developed by Rayleigh and Jeans. The latter had described a blackbody as a system of oscillators of every possible frequency but had failed to obtain a proper description of the observed intensity distribution by applying Maxwell-Boltzmann statistics under the assumption that all energy values were equally accessible to each such oscillator. Planck was able to obtain the observed result by assuming instead that only energy values corresponding to integral multiples of a fixed quantum of magnitude $h\nu$ were available to an oscillator with frequency $\nu$. The key theoretical result is obtained with the calculated mean value $<E>_\nu$ of the oscillator's energy based on this assumption, as given in terms of the ratio of sums given below:

$$<E>_\nu = \sum_{n=0}^{\infty} nh\nu \, \exp(nh\nu / kT) \, / \, \sum_{n=0}^{\infty} \exp(-nh\nu / kT), \qquad (5)$$

where n is an integer, k is Boltzmann's constant and T is the absolute temperature. The classical theory had failed only because integrals over a continuous range of non-integral n were employed instead of the discrete sums in eq. (5).

The key point of interest in the present context is that *the n=0 term in the above sums must be retained* to provide for an accurate representation of the observed spectral intensity distribution. This term does not alter the sum in the numerator of eq. (5), but it makes a decisive contribution to that in the denominator (partition function). According to the theory of statistical mechanics, each term in he above sums corresponds to an allowed state for the system, in this case a photon with energy $E_n = nh\nu$. The zero-energy (n=0) photon is thus an essential ingredient in Planck's long-accepted solution to the blackbody problem. Moreover, as the lowest-energy state available to a photon associated with an oscillator of any given frequency $\nu$, it is also the most frequently populated according to the Boltzmann exponential law, and this at any temperature T. In order to obtain the total intensity distribution it is necessary to integrate over all frequencies from null upwards. It is important to note, however, that zero-energy photon states are present in the distribution for *each value of v*. This situation is illustrated in Fig. 1, in which the various frequencies are represented by the spokes of a wheel. The allowed states for an oscillator of given $\nu$ can be thought of as being plotted as points along the corresponding spoke at a distance from the center of the wheel that is proportional to their energy. Especially if the Boltzmann populations are taken into account, it is found that by far the largest concentration of photons is at the center of the wheel, *i.e.* with exactly zero energy and momentum.

With the introduction of Bose-Einstein statistics, it became possible to obtain a new derivation of the blackbody radiation formula that recognized the indistinguishability of photons [9]. The corresponding distribution function is of the form $(e^{\alpha}e^{h\nu/kT} - 1)^{-1}$ for a photon with energy $E = h\nu$. In order to obtain agreement with experiment, it is no longer necessary to assume that the Planck series of allowed energy values, $E=nh\nu$, exists but rather just the single value for n=1. To obtain the same result for $<E>_\nu$ as in eq. (5), however, it is necessary to set $\alpha=0$ in the Bose-Einstein distribution function, which means that for $\nu = 0$ photons an infinite value results. One is therefore again confronted with several possibilities regarding the interpretation of the statistical mechanical results. Conventionally, one has assumed on the basis of the creation and annihilation of matter hypothesis that zero-energy photons simply do not exist and therefore that the theoretical population is not to be taken literally. The arguments of the preceding section, however, have pointed out logical difficulties with the premise that photons of this limiting energy are not present when those with values infinitesimally close to it can be shown to exist by virtue of the Doppler effect. On this basis it would seem permissible and even more consistent to just assume that the high density of $E = 0$ photons indicated by the theory can be taken at face value. Put the other way around, it would be a very damaging piece of evidence against the massless photon hypothesis if states of zero energy had to be *excluded* from the allowed set in order to achieve a satisfactory representation of the experimental observations, but this is not the case.

Recognition of the ubiquitous presence of massless photons provides a ready explanation for the appearance of electromagnetic radiation in conventional atomic and molecular emission processes that does not involve the creation and annihilation of matter hypothesis. When a system undergoes a spontaneous transition to a state of greater stability, the lost energy must be taken up by another system in its immediate area. The high density of massless photons makes it extremely probable that one of them will be the interacting partner in this exchange, especially because of their known electromagnetic properties. The prototype emission (and absorption) reactions of eqs. (1) and (2) can therefore be rewritten as:

emission: A* + γ (E = 0) → A + γ (E = hν)         (1')

absorption: A + γ (E = hν) → A* + γ (E = 0).       (2')

The new equations differ from the original ones first and foremost in that they are balanced in terms of the number and types of particles that appear on both sides, similarly as one demands for ordinary chemical processes. From the standpoint of energy and momentum conservation, there is no difference between the two pairs of equations because the additional particle in eqs. (1') and (2') carries neither energy nor momentum. Otherwise, the situation is very similar to that for chemical reactions in solution. In cases for which the solvent molecules do not become chemically bound to any of the reactants or products, it is customary to write the corresponding equation without including any such molecules, even though it is clear that they play at least a passive role in the overall process. Alternatively, one includes just enough solvent molecules to ensure particle balance on each side of the reactive equation. In the latter instance it is assumed that the free solvent molecules indicated on one side are identical with those appearing as bound species on the other. In other words, no allowance is made for the creation and annihilation of solvent molecules in chemical equations, and the above arguments show that there is no need to treat photons any differently in the processes of eqs. (1') and (2') in which electromagnetic radiation of non-zero frequency appears as either a product or a reactant.

Even when neither the products nor the reactants of a reaction occurring in solution are conventionally thought of in terms of bound solvent molecules, there are nonetheless often measurable effects attributable to the presence of their unbound counterparts which show up under careful examination of the energetics of the process. The hydrogen bonding phenomenon is a prime example of this type. There is a clear analogy for this type of interaction in the case of photons as well, as exemplified by the radiative corrections in quantum electrodynamics [10-12] that are responsible for the Lamb-Retherford shift [13] and the anomalous magnetic moment of the electron [14]. When employing the hypothesis of photon creation and annihilation, it is necessary to introduce a vacuum field to explain these effects because of the belief that real photons are completely absent in the immediate surroundings of the affected systems. It is a straightforward matter in an alternative interpretation to replace the vacuum field by a sea of massless photons. Indeed, once a high density of such undetectable particles is introduced into the physical model, there would be ample cause for skepticism if effects of this sort were not observed for otherwise isolated atomic and molecular systems. The history of explaining effects which appear to arise out of a void predates the discovery of quantum mechanics with the attribution of mechanical quantities such as linear and angular momenta to electromagnetic fields in classical theory. The presence of a high density of massless photons in the (apparently) empty space separating two conductors again provides a ready explanation for the observed phenomena that does not require the introduction of any new concepts.

In this connection it is well to examine the role of virtual particles in a physical model that recognises the existence of a high density of massless photons in the neighborhood of a given interaction. The term "virtual" generally implies a violation of some conservation law for the particle in question that becomes allowed momentarily by virtue of the Heisenberg uncertainty relation. The need to introduce virtual particles at all into the theoretical treatment, however, is invariably tied up with the belief that it is not possible to write down a balanced equation for the process in question, and thus is closely connected with the creation and annihilation of matter hypothesis. Once one insists that the same particles must always be present throughout the entire course of a given interaction, however, there is another convenient language for discussing such effects, namely that of configuration interaction. For example, one improves upon the single-configuration $1s^2 2s^2$ description of the ground electronic state of the Be atom by combining this term with one of $1s^2 2p^2$ occupation. It is often useful to think of the system as spending a portion of the time in the higher-energy $1s^2 2p^2$ configuration, but this eventuality should not disguise the fact that the wave function as a whole represents a single stationary state. For processes conventionally described in terms of virtual photons, the requirement that configuration interaction terms must all correspond to the same number and type of particles can be met by simply including real photons in various stages of excitation in each of the wave function terms, including those of the massless variety for the dominant terms representing the unperturbed atomic or molecular system.

## IV. Quantum Conditions for Photon Interactions

The quantum jumps associated with photon interactions provided an important clue regarding the particle nature of light. In his exposition of the photoelectric effect [4] Einstein reversed a trend away from the Newtonian view [15] of light as "corpuscles". He showed that surface ionisation of metals could be most consistently explained by assuming that a single quantum of light gives up all its energy to a single electron. He used the word "heuristic" in describing his ideas [4] because the (exclusively) wave theory of electromagnetic radiation was widely accepted by the physics community at that time. While there can be general agreement that the photoelectric effect is inconsistent with a totally wave-like nature for light, it still must be regarded as extraordinary that any particle would transmit *all* its translational energy to a single electron in a given interaction. Such a property of photons is consistent with the concept of annihilation, because it is necessary to assume that a particle which has gone out of existence does so by leaving behind all its energy and momentum. Once it is assumed instead that the photon does not go out of existence as a result of the photoionisation, however, but rather assumes a massless state that as a consequence of the Bohr frequency relation [1] defies direct experimental observation, it becomes necessary to look more closely at the dynamics of this process to better understand the nature of the quantisation phenomenon.

To this end it is instructive to apply the laws of energy and momentum conservation to the radiative absorption process, as depicted in Fig. 2. If a photon ? were to give off an *arbitrary* amount $\Delta E_\gamma$ of its total energy $E_\gamma = h\nu$ to an atom A with mass $M_A$, its momentum would decrease by $\Delta p_\gamma = \Delta E_\gamma/c$. If the atom were to remain in the same internal state, this amount would appear exclusively in the form of translational energy, which means that the momentum of the atom would change by $\Delta p_A = (2M_A \Delta E_\gamma)^{1/2}$. Conservation of momentum requires that $\Delta p_A$ and $\Delta p_\gamma$ be equal. For small $\Delta E_\gamma$ this can never be the case, however, in view of the large mass of A. Setting $\Delta p_A$ equal to $\Delta p_\gamma$ shows that $\Delta E_\gamma$ would have to be equal to twice the rest energy of A or $2M_A c^2$, which corresponds to the GeV range [16]. There is a solution to this dilemma, however, namely to have a part of the photon's energy be added to the internal energy of the atom, *i.e.* to reach another electronic state of the more massive system. If the excited electronic state differs by $h\nu'$ in energy from that of the initial state, conservation of momentum then requires that

$$\Delta p_\gamma = \Delta E_\gamma/c = \Delta p_A = [2M_A (\Delta E_\gamma - h\nu')]^{1/2} \tag{6}$$

which is possible provided $h\nu'$ is only slightly smaller than $\Delta E$, again by virtue of the relatively large mass of A as well as the magnitude of c.

It is important to distinguish between two aspects of the absorption process in the foregoing discussion. First, the quantised nature of the atomic spectrum is connected directly with the large disparity between the respective masses of the atom and the photon. When one considers the translational motion of the atom, it is recognised that the energy levels available to it are actually continuous. It is the requirement of momentum conservation that restricts the possible transitions between different states of the same atom and thereby produces the quantisation phenomenon. On the other hand, on the basis of these arguments by themselves there is no restriction put on the magnitude $\Delta E_\gamma$ of the energy lost by a photon in the absorption process, save that it be less than or equal to its total energy, $E_\gamma = h\nu$. Indeed, the analogous excitation brought about by electron impact is well known [17]. One is thus still left with the conclusion that there is something special about a zero-energy, zero-momentum state of the photon, even though many aspects of the absorption phenomenon can be explained by just assuming that the photon is a particle of relatively small mass compared to the system with which it interacts.

The fact that the energy transferred in the above process is exactly equal to $E_\gamma = h\nu$ thus still requires explanation. In other words, why doesn't a photon give off only part of its energy in inducing a transition in another system? Dirac used time-dependent perturbation theory [18] to answer this question, arguing that the incident radiation introduces a frequency-dependent term in the Hamiltonian of the atomic system. A resonance condition results according to which the energy of the most probable atomic transition, $h\nu = E_i - E_f$ must be the same as the energy of the incident photon, $E\gamma$.

The prospect of a massless photon being formed as a result of this energy exchange (rather than that the original photon is annihilated in the process) suggests a somewhat different interpretation for this phenomenon, however, one that does not rely on the assumption of wavelike properties for the incident radiation. If one simply looks upon the process as a collision between an atom and a photon moving with speed c, it seems plausible to demand that the observed energy exchange take place over a relatively small but finite period of time. As a consequence, the temporal requirements of the interaction are far more readily fulfilled by an outgoing system *whose velocity has been considerably reduced below the speed of light* in a vacuum. As long as the departing photon possesses a non-zero amount of energy, this condition can never be fulfilled, but as has been pointed out in Sect. II, a *massless* photon is free of any such restriction, and thus can move at any speed less than c, including zero. In this view, the only practical means available to a photon to reduce its energy by virtue of an atomic collision is to assume a massless state, so that its relative speed compared to the system with which it interacts can be made as close to zero as possible. Accordingly, this interaction mode *represents the only inelastic process available to a system of zero rest mass,* since it is otherwise forced to move with the speed of light away from the region of interaction if it possesses any non-zero amount of translational energy.

By combining this result with the conservation of energy and momentum arguments first discussed, it is seen that the quantum characteristic associated with radiative absorption (and emission [19]) can be deduced exclusively on the basis of the widely different rest masses of the photon and the system with which it interacts, with no need to give separate consideration to the wave characteristics of the associated electromagnetic field. Rather one is led to conclude from knowledge of the internal energy states of the interacting system and the magnitude of its rest mass exactly which photon energy is required to induce maximum transition probability. The magnitude of this transition probability itself cannot be determined quantitatively on the basis of the above information alone, and thus for this purpose one does have to introduce some additional information about the nature of the perturbing Hamiltonian. This state of affairs does not affect the main conclusion in the present discussion, however, namely that the properties expected on the basis of the special theory of relativity for a massless but nonetheless existent system are sufficient in themselves to allow for a suitable explanation of the observed tendency of photons to give up all their translational energy upon interacting with other particles.

## V. Conclusion

The preceding discussion has examined the question of whether photons can exist with exactly zero energy. The creation and annihilation of matter as a principle denies that this is possible, asserting instead that a photon must always move with the speed of light relative to any observer and thus have $E > 0$ by virtue of the mass dilation formula of the theory of special relativity. Examination of the latter expression shows, however, that a particle of zero rest mass can move with any speed down to and including v = 0 when the E = 0 condition is present. A key point that also needs to be considered in this context is that according to the Bohr relation a photon beam with E = 0 must possess a null frequency and infinite wavelength and as a result could never be detected by direct experimental means. There is thus a clear alternative to the conclusion that E = 0 photons do not exist, namely that it is simply impossible to detect them when they occur. More generally it must be recognized that there can never be any objective proof that any particle has either been created or annihilated because one can never be certain that what appears to be non-existent is not in fact present in exactly such an undetectable state as indicated by the Bohr frequency relation for E = 0 photons.

There is also *positive evidence* that photons with zero energy do exist, however, and that is obtained from two sources: the relativistic Doppler effect and blackbody radiation theory. In the former case it can be seen that photons with energies only infinitesimally greater than zero must exist because according to the Doppler effect it is possible to bring the frequency of this radiation into an easily detectable range by having the observer move with a speed close to c toward the source. It would be incongruous to claim that the existence of the photons was conditional upon the state of motion of the observer relative to them. But since photons can exist with energies only infinitesimally greater than zero, one must conclude that they also exist at the limit of exactly zero energy as well. By analogy to the limit-taking procedure in mathematics, it is possible to formulate a physical principle of infinitesimally extended validity for theories which cannot be explicitly tested at special points just outside their range of experimental verifiability. In other words, a physical theory which is known to hold infinitesimally close to a certain point in its domain must also be valid at the limit itself, just as the impossibility of directly evaluating the ratio of sin x/x at x = 0 does not prevent us from assigning a unique limiting value of unity to this quantity.

This conclusion brings photons into full equivalence with other particles of non-zero rest mass since it means that kinetic energy can also be removed from them on a continuous basis until a minimal value of zero has been reached. The same principle employed above to come to this realization also enables one to prove that the relative speed of E = 0 photons can lie well below v = c, including v = 0. In this case the mass dilation formula is at issue, and since it is known to correctly predict that v = c for photons of any non-zero energy/frequency, including one infinitesimally close to a zero value, it must also be concluded that its predictions for the speed of photons in the limiting case of E = 0 are also valid. On this basis one can call into question the conventional view that photons can never be at rest with respect to any observer. One must only forego any hope of detecting photons in this translational state because of the impossibility of measuring an infinitely long wavelength.

Moreover, the same principle allows us to determine the population density of E=0 photons, this time by virtue of blackbody radiation theory. The energy distribution proposed for each oscillator in the blackbody cavity restricts photons to quantised energy values proportional to the fundamental frequency of the corresponding standing wave pattern. The theoretical populations of non-zero energy states are experimentally verifiable for any frequency down to but not including a zero value. The corresponding theory, both in Planck's original formulation based on the Maxwell-Boltzmann distribution and in the later treatment employing Bose-Einstein statistics, indicates that the population of the E = 0 photon levels is inexhaustibly large. Since with the aid of the Doppler effect it can be shown that the same theory correctly predicts the population of photon states corresponding to energies that are infinitesimally close to zero, it is inconsistent on the basis of the above principle to argue that the analogous predictions for photons of exactly zero energy are not to be taken literally.

As a consequence the traditional/view that photons do not exist prior to radiative emission processes or subsequent to the corresponding absorption processes is also shown to be unfounded. One can explain such phenomena quite consistently in terms of the existence of a high density of zero-energy photons in all locations of the universe, exactly as indicated by blackbody radiation theory. It is also possible to equate the traditional concept of a vacuum field, so critical to the theory of quantum electrodynamics, to the existence of real photons corresponding to a null (and therefore undetectable) frequency.

The quantised nature of photon interactions emerges in a straightforward manner in terms of the existence of photons with E = 0. In absorption the energy lost by an incident photon c an only be taken up partially as translational energy for the heavier atomic system in order to satisfy the momentum conservation law. The greatest part of the photon's energy must be used to change the interacting system's internal state, which circumstance greatly restricts the values of ΔE for which the photon's energy can be absorbed and therefore causes the quantisation phenomenon to occur. The fact that the final state of the absorbed photon has exactly zero energy arises because the entire process must take place with both particles remaining in a small volume for at least a short period of time. This requirement cannot be met by a photon with E ≠ 0 because it must always move with the speed of light away from the region of interaction. Photons with exactly zero energy can move with speeds considerably less than v = c, however, by virtue of the mass dilation formula, and hence the absorbing photons must lose all their energy to satisfy this requirement. In the last analysis one can relate this situation to practical experience. If two people want to have a business transaction and one of them is on a speeding train while the other is waiting at the station platform, it is first necessary that they come to rest with respect to each other. But a photon of non-zero energy can never be at rest with respect to an observer, hence the only practical way of interacting with a second (massive) particle is for it to give up all its energy so that a speed less than v = c is possible.

There are more general ramifications to the present study, however, which go beyond the realm of photon interactions. If it is possible to understand these phenomena without requiring the creation or annihilation of photons, then there is good reason to reexamine other processes that are traditionally interpreted in terms of particles passing to and from existence. It may be necessary to return to the seemingly plausible view of nature as dramatized in the words of Lucretius [21]: "Nothing can be created from nothing" (or in the equivalent remarks of Shakespeare in *King Lear:* "Nothing will come of nothing"). In this view, the creation and annihilation of material particles can only be a heuristic explanation for submicroscopic phenomena because the true building blocks of nature are indestructible. Again in the words of Lucretius [22], who was mirroring the thoughts of Democritus in his atomic theory proposed 500 years earlier: "Material objects are of two kinds, atoms and compounds of atoms. The atoms themselves cannot be swamped by any force, for they are preserved indefinitely by their absolute solidity." On this basis one could return to the time-honored approach of describing all interactions in terms of balanced equations, such as has been done for the absorption and emission of electromagnetic radiation in the present eqs. (1',2'). Such an approach would not require changes in the computational methods of quantum field theories, rather only in their interpretation in terms of real particles that are undetectable by virtue of their absolute lack of energy/mass. In this way it is no longer necessary to view results such as the explicitly defined populations of E = 0 photon states in blackbody radiation theory as mere theoretical artefacts. Instead, a substantial unification of the overall theory of such processes is achieved which leaves open the distinct possibility that all particles are just as indestructible (and uncreatable) as Lucretius claimed nearly twenty centuries earlier.

**Acknowledgment**

The author wishes to express his sincere gratitude to Dr. Gerhard Hirsch for numerous helpful discussions during the course of this work and for his critical reading of the manuscript. The financial support of the Deutsche Forschungsgemeinschaft in the form of a Forschergruppe grant is also hereby gratefully acknowledged.

**Figure Captions**

Fig. 1:  Schematic diagram representing the distribution of allowed energy levels of oscillators in Planck's theory of blackbody radiation. Each spoke of the wheel corresponds to a fundamental frequency $v$, whose energy quantum, $E = hv$, is always proportional to the distance between adjacent points on such a radius. Each such (equally spaced) point thus corresponds to an allowed energy level, one of which is always found at the hub of the wheel for each spoke, *i.e.*, $E = 0$ is allowed for every value $v$. The magnitude of the energy quantum is shown to decrease monotonically as one proceeds in a clockwise fashion from the twelve o'clock position. The partition function used in Einstein's explanation of the blackbody radiation phenomenon must include the $E = 0$ levels explicitly for each fundamental frequency in order to obtain results which agree with experiment. It is thus clear from the diagram that the highest concentration of allowed energy levels by far is located at the hub of the wheel, and the form of the Boltzmann exponential factors, $\exp(-E/kT)$, insures that the highest photon population always occurs for this energy value.

Fig. 2:  Energy level diagram detailing the role of conservation laws in determining whether a given radiative absorption process is allowed or not. At the top of the diagram, the system is to retain the same internal energy $E_S''$ in the transition, *i.e.* the two levels shown differ only in translational energy $\Delta T_S = \frac{1}{2} m_S \overline{\Delta p_S^2}$ (nonrelativistic theory), where $m_S$ is the mass of the system and $\Delta p_S$ is the corresponding change in the momentum of its center of mass.

Such a radiative process is forbidden by the law of conservation of linear momentum, because the rest mass of the photon is so much smaller than that of the system ($\Delta p_S > \Delta p_\gamma$). Radiative absorption can occur, however, if the system changes its internal (from $E_S''$ to $E_S'$) as well as its translational energy, as depicted in the lower part of the diagram. Under these circumstances the momentum conservation law can be satisfied for a particular value of the $\Delta p_S$, namely one that is equal to $(E'_S - E''_S + \overline{\Delta p_S^2}/2 m_S)/c$, where c is the speed of light. This condition rules out the occurrence of a radiative absorption process in which the system's translational energy does not change at all, also as indicated. Thus the "quantised" nature of radiative transitions is intimately connected with the photon's vanishing rest mass.

Fig.1

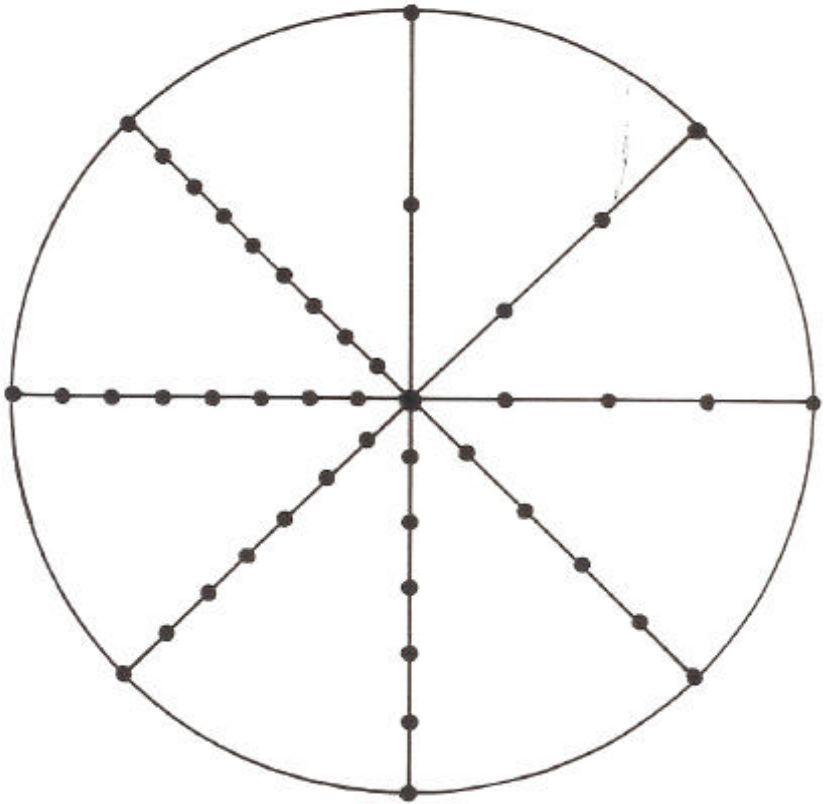

Fig. 2

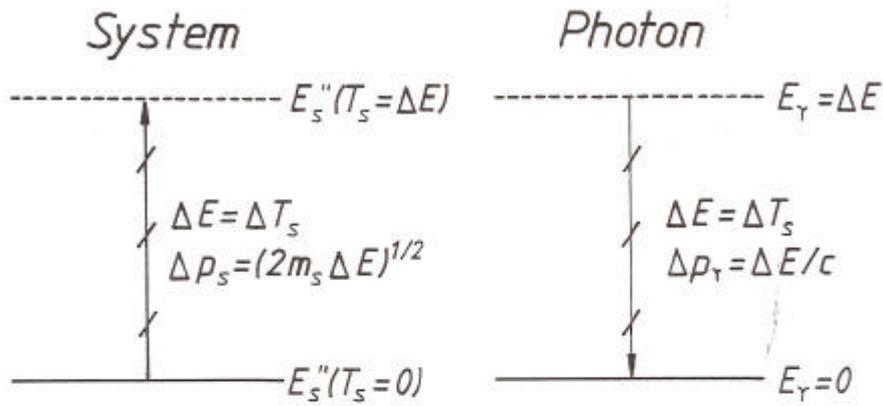

Forbidden: $\Delta p_s > \Delta p_\gamma$

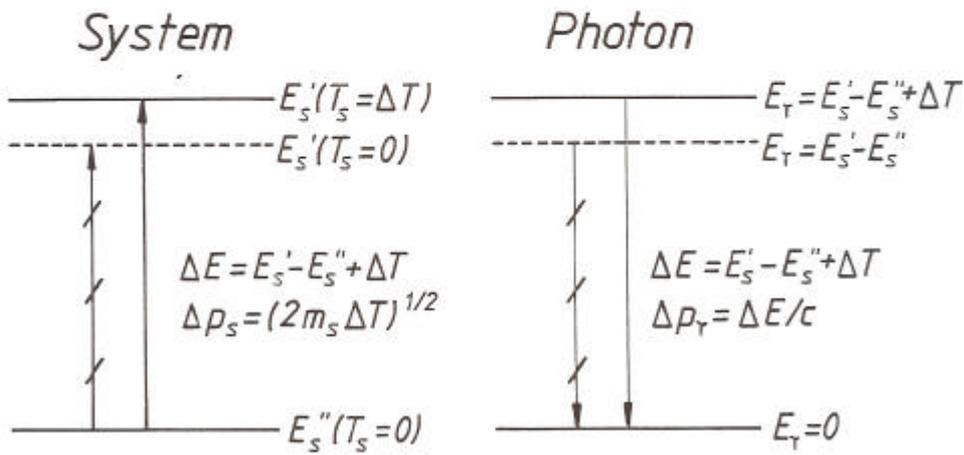

Allowed: $\Delta p_s = \Delta p_\gamma = [E_s' - E_s'' + \overline{\Delta p_s}^2/2m_s]/c$